\documentclass[conference]{IEEEtran}
\usepackage{multirow}
\usepackage[caption=false]{subfig}

\usepackage{cite}

\ifCLASSINFOpdf
  \usepackage[pdftex]{graphicx}
\else
\fi
\usepackage{epstopdf}
\usepackage{adjustbox,lipsum}

\usepackage{array}
\usepackage{url}

\usepackage{todonotes}

\begin{document}
%
\title{Using Changeset Descriptions as a Data Source to Assist Feature Location}

\author{\IEEEauthorblockN{Muslim Chochlov, Michael English, Jim Buckley}
\IEEEauthorblockA{Department of Computer Science and Information Systems\\
University of Limerick\\
Limerick, Ireland\\
Muslim.Chochlov@lero.ie,
Michael.English@ul.ie,
Jim.Buckley@lero.ie}}


%


\maketitle

\begin{abstract}
Feature location attempts to assist developers in discovering functionality in source code. Many textual feature location techniques utilize information retrieval and rely on comments and identifiers of source code to describe software entities. An interesting alternative would be to employ the changeset descriptions of the code altered in that changeset as a data source to describe such software entities. To investigate this we implement a technique utilizing changeset descriptions and conduct an empirical study to observe this technique's overall performance. Moreover, we study how the granularity (i.e. file or method level of software entities) and changeset range inclusion (i.e. most recent or all historical changesets) affect such an approach. The results of a preliminary study with Rhino and Mylyn.Tasks systems suggest that the approach could lead to a potentially efficient feature location technique. They also suggest that it is advantageous in terms of the effort to configure the technique at method level granularity and that older changesets from older systems may reduce the effectiveness of the technique. 
\end{abstract}

\begin{IEEEkeywords}
Feature location, information retrieval, software repositories
\end{IEEEkeywords}
%
\IEEEpeerreviewmaketitle

\section*{Note to Practitioners}
This is the accepted version of the paper submitted to the 2015, IEEE 15th International Working Conference on Source Code Analysis and Manipulation (SCAM). The final version should be accessible at \url{https://doi.org/10.1109/SCAM.2015.7335401}.

\section{Introduction}
\label{sec:introduction}
Independent studies suggest that software maintenance constitutes over 60\% of the software development budget \cite{Brooks1995,Canfora2001}. Of all software maintenance tasks, enhancements account for over 75\%, while corrective activities are less frequently performed\cite{Lientz1980,Canfora2001}. \emph{Program comprehension} is inevitable when engaging in software maintenance tasks\cite{Rajlich2002}. Developers approach the problem of program comprehension by studying and iteratively expanding their knowledge about a software system \cite{Mayrhauser1995}. Different strategies are utilized by developers, depending on their knowledge of a system, their domain expertise, their programming language experience, and their personal preferences\cite{Pennington1987,Soloway1984,Littman1986,Shaft2006,OBrien2004}.

Various case studies confirm, that over 50\% of developers' maintenance effort is spent on understanding the source code\cite{Fjeldstad1983, Canfora2001}. Given the effort associated with software maintenance, program comprehension can be considered one of the most expensive activities in software development. 

Hence there is an obvious need for movement towards the automation of program comprehension. Biggerstaff et al. formulated the \emph{concept assignment problem} as a discovery and mapping of application domain concepts to their source code counterparts to help programmers\cite{Biggerstaff1994}. Authors claim that deriving a plausible, or correct concept assignment is a challenging task because of a significant gap between application and code domains\cite{Jordan2015}. Instead, a strategy often employed by developers is to search for \emph{clues} in source code until enough evidence is obtained to assign a concept\cite{Soloway1984}. 

Automated program comprehension techniques such as \emph{feature location} (FL) emerged to assist developers. Though no standard definition of a \emph{feature} exists, a common view is that a feature represents a functional requirement\cite{Wilde1995}. The term \emph{concept} is usually used as a synonym of a feature\cite{Dit2011}, or to refer to a superset of features\cite{Scanniello2011}. According to Dit et al. a \emph{feature location technique} (FLT) is expected to return an \emph{entry point} leading to the implementation of the feature under inspection as part of its output\cite{Dit2011}. FLTs are usually classified into dynamic and static, where static could be further divided into structural and textual \cite{Dit2011,Rubin2013}. Dynamic FLTs utilize execution traces of a program, while static approaches leverage the lexical and control or data flow information within source code. \emph{Information retrieval} (IR) is commonly applied with textual FLTs to leverage the lexical information available in source code\cite{Dit2011}. Source code comments and identifiers are usually used to describe software entities, but more recently several FLTs have attempted to take advantage of additional historical information obtained from third party tools such as \emph{version control systems} (VCS) or \emph{issue tracking systems} (ITS) in their aim to identify features\cite{Chen2001,Cubranic2005,Ratanotayanon2010,Zamani2014,Kevic2014}.

VCSs store a set of changesets, where each changeset contains a textual description written at a high level of abstraction that refers to the rationale for source code changes touched by this changeset. Such characteristics make changeset descriptions a reasonable alternative or a supplement to traditionally used textual data sources (i.e. comments and identifiers). In this paper we present a FLT that Aggregates Changeset descriptions to annotate relevant software artifacts for IR (ACIR).

We implement ACIR and perform a preliminary assessment of the efficiency of this approach, comparing it to the results to similar existing FLTs utilizing comments and identifiers. The granularity at which software artifacts are to be indexed and later retrieved is one interesting characteristic of this approach that should be assessed. Unlike local comments and identifiers, changesets are known to touch several software artifacts in source code. Therefore, artifacts of coarse and fine granularity will share more common terms, which may in turn affect IR results. We experiment with file level and method level granularities, which seem to be favoured by the vast majority of FLTs\cite{Dit2011}. Likewise, lines of code on average have more than one changeset applying to them \cite{Chen2001}. However, it is not clear if including all the changesets' descriptions will improve the efficiency of ACIR. By including all the changesets we will expand the dataset available to the IR engine, that could potentially lead to improved efficiency. On the other hand, older changesets might be overwritten for a reason and including them will introduce undesirable noise. We perform our evaluation of the overall performance and these two characteristics using two open source projects \emph{Rhino}\footnote{\url{https://developer.mozilla.org/en-US/docs/Mozilla/Projects/Rhino}} and \emph{Mylyn.Tasks}\footnote{\url{https://eclipse.org/mylyn/}}.

\section{Related Work}
\label{sec:related_work}

Historical information has been long recognized as a valuable data source for various software maintenance activities \cite{Zimmermann2005,Mockus2000}. Particularly in program comprehension, historical data has been found useful for \emph{traceability link recovery} \cite{Ali2013a,Mazzeo2013,Ali2013}, \emph{impact analysis} \cite{Zanjani2014}, and \emph{bug localization} \cite{Sisman2012,Wang2014}. In contrast limited FLTs have been developed that leverage historical data. Of more than 60 unique FLTs identified by Dit et al.\cite{Dit2011}, only 3 utilized historical data \cite{Chen2001,Cubranic2005,Ratanotayanon2010}. We subsequently reviewed the literature of FLTs from the year of 2010 to 2014 and found 2 more FLTs utilizing historical data\cite{Kevic2014,Zamani2014}.

According to the taxonomy by Dit et al. the IR approaches tend to prevail in the FLT research literature that rely on text analysis\cite{Dit2011}. The historical techniques analysed in this section mostly try to locate features employing various IR models. Chen, Ratanotayanon, and Zamani indexed and searched documents using \emph{vector space model} (VSM), and Cubranic utilized \emph{latent semantic indexing} (LSI). VSM and LSI both belong to the family of algebraic IR models, where term statistics are important to the model's efficiency and \emph{the term frequency - inverse document frequency} (TF*IDF) function helps calculate a term's importance \cite{SaltonGerardandMcGill1983}. Chen, Cubranic, and Ratanotayanon utilized a generic TF*IDF formula, whereas Zamani et al. adjusted TF*IDF to not only reflect term frequency across documents but also to include date into its calculation, thus altering terms' scores. The noteworthy exception is the technique by Kevic and Fritz that implements a custom dictionary model instead of using the IR model to address differences in source code and change request vocabularies\cite{Kevic2014}. Yet, the authors incorporated TF*IDF scoring into their approach. The input to all the techniques is a search query typed by user or extracted from change request, and the output is organized as a ranked list of documents. The granularity of the techniques was almost equally represented by either file or method level selection.

The type of information available in historical data largely depends on the source this data was extracted from. VCSs appear to be the most widely considered source of historical data and were used by all historical FLTs\cite{Chen2001,Cubranic2005,Ratanotayanon2010,Zamani2014,Kevic2014}. VCSs have rich evolutionary information (i.e. where both textual data and metadata are easily accessible) and are broadly available. The second popular source of data were  ITSs. They were used by all authors except Chen et al. Less frequently used historical tools included the task context tool Mylyn utilized by Kevic and Fritz and communication applications such as \emph{email}, utilized by Cubranic and Murphy. Though similar sets of historical tools was used by all techniques, their actual application varied greatly depending on the heuristics employed. 

The technique proposed by Ratanotayanon et al. treated VCS changesets as documents for IR, with their metadata filling appropriate search fields (e.g. author, timestamp)\cite{Ratanotayanon2010}. Where possible, matching textual issue descriptions from ITS were used to augment such documents with additional lexical information. The resulting ranked documents (i.e. changesets) would point to source code artifacts, and the \emph{abstract dependency graph} of such artifacts could be used to further expand the list of relevant results. Techniques by Kevic and Fritz, and Cubranic share a somewhat similar approach, where a new change request is first matched against the collection of issues from ITS. It is then a problem of finding relevant VCS changeset to link the change request to source code. The technique by Zamani2014 utilized source code comments and identifiers as a source of data to be used in IR\cite{Zamani2014}. The authors adjust the scoring function of the IR model with timestamp data available from VCS to favour the terms of comments and identifiers altered by the most recent changesets.

The generally observed tendency is that textual information from historical tools and its subsequent processing by IR is common in such FLTs. However, there are approaches that experiment with other historical metadata, such as timestamps, to assist FL. In the scope of the FL problems stated in section \ref{sec:introduction} the technique by Chen et al. looks as the most relevant to our approach\cite{Chen2001}. Chen et al. utilized changeset descriptions of VCS to annotate source code lines for textual search. For each line of code all associated changeset descriptions were aggregated and matched against search query. The source code file would then be ranked according to the number of matching lines. Therefore, text from changeset descriptions served as an alternative to comments and identifiers, which are usually used in textual FLTs. The biggest difference between our approach and that of Chen et al. is how we look at the process of FL. Chen et al. were interested in a \emph{grep}\footnote{\url{http://www.gnu.org/software/grep/manual/}} like code search approach, where changeset descriptions would substitute source code lexical data. Instead, we treat source code as a collection of artifacts of a given granularity (like many textual FLTs utilizing comments and identifiers do\cite{Marcus2004a}), where changeset descriptions are aggregated to describe an artifact.

\section{The Technique}
\label{sec:technique}
\begin{figure*}[!t]
\centering
\includegraphics[width=0.75\textwidth]{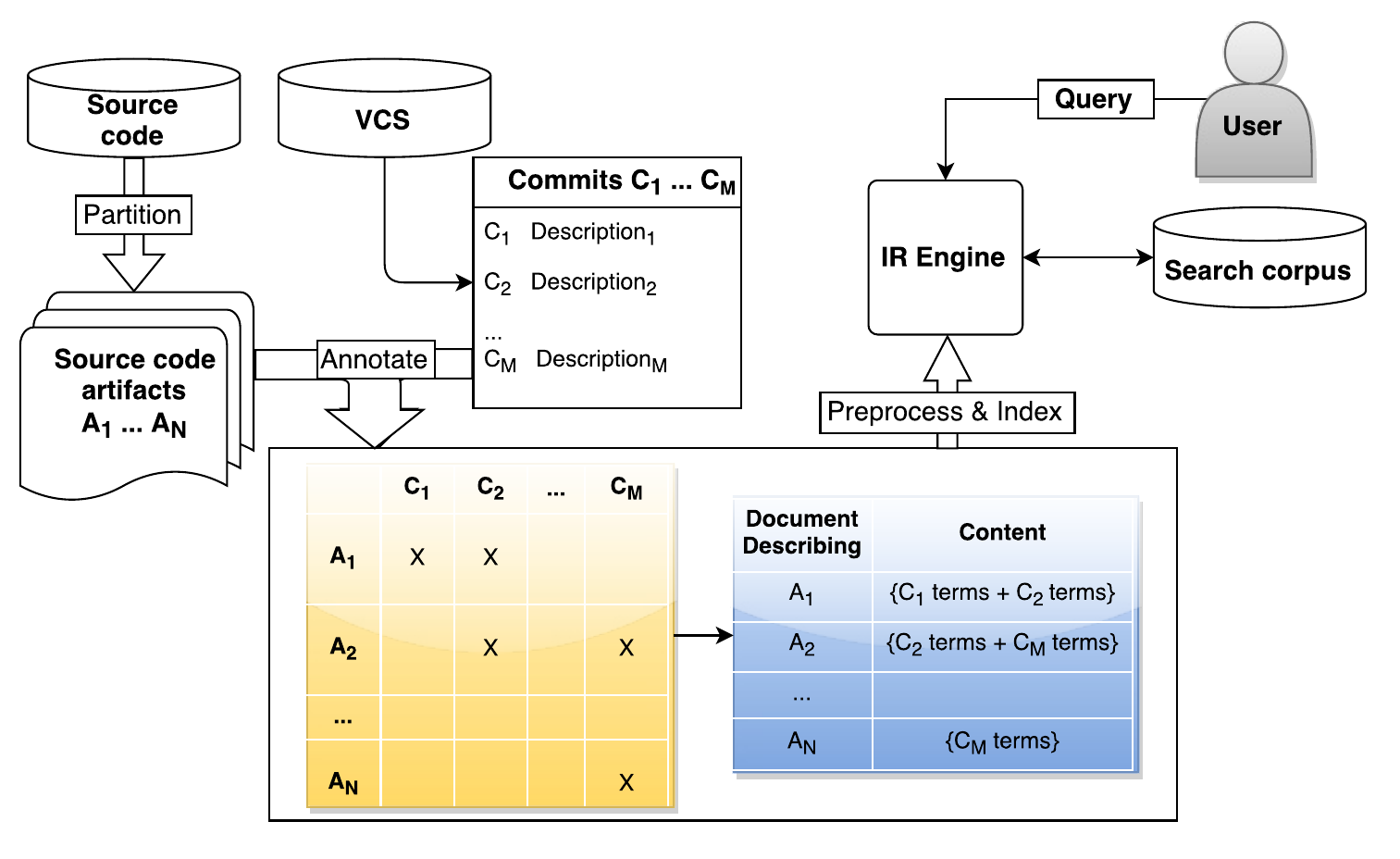}
\caption{The design of ACIR}
\label{fig:technique_png}
\end{figure*}

In this section we present the ACIR technique. The technique first partitions source code into artifacts of a given granularity. For each source code artifact it then retrieves relevant changeset descriptions from VCS to collectively annotate the entire artifact. IR leverages the resulting documents (i.e. software artifacts with aggregated changeset descriptions as their content) in a search corpus retrieving them when a search query is submitted. The technique allows for different levels of granularity to be chosen. It is also capable of annotating the artifacts with the most recent changeset descriptions or with all changeset descriptions available for this artifact. The schema in Figure~\ref{fig:technique_png} highlights essential components of the technique. In general it follows steps described by Marcus et al. that are common to many textual FLTs with the exception of the second step\cite{Marcus2004a}. We implemented the technique in the \emph{Java} programming language with the help of such third party libraries as \emph{JavaParser}\footnote{https://code.google.com/p/javaparser/}, \emph{JGit}\footnote{https://eclipse.org/jgit/}, and \emph{Apache Lucene}\footnote{http://lucene.apache.org/core/}. The design of this approach does not impose limitations on the VCS being used or the programming language of a software system's source code. However, in this work we used libraries that allow for interaction with \emph{Git}\footnote{http://git-scm.com/} VCS and to parse source code files written in Java programming language.

As seen from Figure~\ref{fig:technique_png} there are 3 major steps that have to be completed before a user can engage in FL.
\begin{enumerate}
\item \textbf{Partition.} First, the source code of a software system at a given revision is partitioned into artifacts of preselected granularity. For this task we rely on JavaParser library, which takes a Java project and iteratively visits source code components. We then obtain a set of source code artifacts along with their names and line numbers. 
\item \textbf{Annotation.} In this step we annotate each source code artifact with corresponding changeset descriptions that collectively describe this artifact. To access changeset records we connect to Git VCS using JGit library. After a connection is established, we take each source artifact from a set and retrieve corresponding changeset information for each line of code belonging to the artifact. Since it is usual for a changeset to describe several lines of code, we maintain a list of already seen elements to keep only unique records. Git supports retrieval of the most recent changesets through \emph{git annotate} command and retrieval of all changesets through \emph{git log -L} command. We rely on these commands when annotating a source code artifact with the most recent or all historical changeset data. At this stage we treat each software artifact as a textual document, where the content of such a document is comprised of the set of terms extracted from descriptions of matching changesets. Such a representation of data is ready to be consumed by the IR engine.
\item \textbf{Preprocessing and Indexing.} Before the source code artifacts are indexed by the IR engine, we pre-process the textual data coming from the commit descriptions for each of them. This includes special character filtering, removal of \emph{stop} words (i.e. the likes of ``a'', ``the'', ``on'' etc.), turning all words into lower case, and applying Porter stem filter\cite{Porter1980}. When text preprocessing is completed, source code artifacts are ready to be consumed by the IR engine, stored in a search corpus, and indexed for fast retrieval. We employ the Apache Lucene framework to assist with IR activities, which by default stores artifacts using VSM and applies a standard TF*IDF function to calculate term scores.
\end{enumerate}

\begin{table*}[!t]
\renewcommand{\arraystretch}{1.3}
\caption{Subject Systems' Statistics}
\label{tbl:subject_stats}
\centering
\begin{tabular}{llrrrrrrrrr}
\hline
\multirow{3}{*}{\begin{tabular}[c]{@{}l@{}}Subject\\ System\end{tabular}} & \multirow{3}{*}{\begin{tabular}[c]{@{}l@{}}Initial\\ release\end{tabular}} & \multirow{3}{*}{\# Files} & \multirow{3}{*}{\# Methods} & \multirow{3}{*}{\begin{tabular}[c]{@{}r@{}}Average\\ methods \\ per file\end{tabular}} & \multirow{3}{*}{LOC} & \multirow{3}{*}{\begin{tabular}[c]{@{}r@{}}\# Commits \\ (master)\end{tabular}} & \multicolumn{4}{c}{\begin{tabular}[c]{@{}c@{}}Average of distinct\\ commits per artifact\end{tabular}} \\ \cline{8-11} 
                                                                          &                                                                            &                           &                             &                                                                                        &                      &                                                                                 & \multicolumn{2}{c}{Function level}                & \multicolumn{2}{c}{File level}                     \\ \cline{8-11} 
                                                                          &                                                                            &                           &                             &                                                                                        &                      &                                                                                 & Recent                   & All                    & Recent                   & All                     \\ \hline
\begin{tabular}[c]{@{}l@{}}Rhino\\ (rev. 06710fa1)\end{tabular}           & Autumn 1997                                                                & 468                       & 6120                        & 13.07                                                                                  & 90435                & 3136                                                                            & 2.11                     & 3.56                   & 9.44                     & 15.44                   \\
\begin{tabular}[c]{@{}l@{}}Mylyn.Tasks\\ (rev. 68400a63)\end{tabular}     & \begin{tabular}[c]{@{}l@{}}June 2005\\ (first public release)\end{tabular} & 1086                      & 8517                        & 7.84                                                                                   & 117348               & 8576                                                                            & 2.27                     & 3.34                   & 11.48                    & 15.54                   \\ \hline
\end{tabular}
\end{table*}

After the source code artifacts are indexed and search corpus is created, a user may start submitting search queries to locate features in source code. A search query text is preprocessed using the same filters as described above in the indexing step. The similarity between a query and documents in search corpus is then calculated, and the ranked list of documents is returned as a result.

\section{Empirical Study}
\label{sec:experiment}

To evaluate ACIR we first implemented it as the tool described in Section~\ref{sec:technique}. To guide this empirical study we formulate research questions in Section~\ref{subsec:rqs}, select open source projects for study in Section~\ref{subsec:subject_sys}, and analyse them with regards to the method and metrics of this study in Section~\ref{subsec:metrics_method}.  

\subsection{Research Questions}
\label{subsec:rqs}
In this paper we propose ACIR, which is a FLT that leverages changeset descriptions of VCS to collectively annotate program artifacts at a given level of granularity. The technique utilizes IR employing VSM to construct and query search corpus. This empirical study's first research question addresses the efficiency of such an approach:

\textbf{RQ1:} How efficient is ACIR?
\newline

The granularity in FL could be roughly divided into coarse and fine grained, where the latter is associated with any component below the file level. Even though Rajlich and Gosavi argue class level granularity is the closest to a concept, Dit et al. claim that the more accurate and specific FLT should exercise the finer level of granularity\cite{Rajlich2004,Dit2011}, and the FLT community have, by-in-large adopted this philosophy \cite{Rajlich2004,Buckley2005,Cleary2008,Scanniello2011,Marcus2004a,Biggers2012}. However, authors of FLTs rarely justify the selection of one level of granularity over the other.
 
We assume that local comments and identifiers that are used to describe methods, are generally unique to those methods and are not shared between other methods of a file. A software artifact of a file level granularity is then described by all comments and identifiers available to the methods of that file. In case of a file level granularity, the document dimensionality of \emph{the document*term} matrix of VSM is significantly reduced compared to method level granularity, whereas the term dimensionality remains unchanged. As a result there are greater chances of increased scalar product between a query and a document at a file level of granularity, which will positively affect cosine similarity making a document to stand out in the search results. Yet, the method level of granularity has its advantage of pointing more accurately to a place in source code. Ultimately, it is the effort required of a developer to reach the correct location in source code that is important and was previously used as a metric by other researchers\cite{Petrenko2013}. In contrast to local method comments, changesets can describe several methods within a file and across the files. As seen from Table~\ref{tbl:subject_stats} software artifacts at a method level granularity are described on average by at least 2 changesets, which means we are likely to have more meaningful terms available. Therefore, the gap between the content of a document at method and file granularity might not be as significant as in case when comments and identifiers are used. However, there is also a possibility that several changesets will describe several methods or files, thereby preventing one of them to stand out. Still, we assume that the increased scores of artifacts at a method level granularity and their ability to accurately point to places in source code might in the end prove to be more advantageous in terms of the effort. We then formulate the second research question as:
 
\textbf{RQ2:} How does granularity of ACIR affect the effort?
\newline

The range of changeset inclusions is unique to historical FLTs and to the best of our knowledge has not been studied in the FL literature. Still, we can draw some interesting conclusions from the existing research presented by Zamani et al. and Chen et al. \cite{Zamani2014,Chen2001}. In the first work the authors make a claim that older terms should receive lower scores when ranking due to the assumption that they become obsolete during the natural lifecycle of a software system (note that, in that work terms were extracted from source code and changesets' date metadata was used to adjust their scores). In contrast, Chen et al. used all available historical changesets equally to annotate each line of code for FL. Both works reported the increased efficiency of their techniques. Though these two approaches are hardly comparable, they lead to an interesting question about the impact that changeset-range has on the effectiveness of ACIR. We focus on the two most extreme points of the changeset range (i.e. most recent changesets opposed to all changesets of an artifact), since they would probably most noticeably highlight any existing differences. These assumptions lead us to the third question:

\textbf{RQ3:} How does including only the most recent, as opposed to all changesets, affects the effectiveness of ACIR?

\subsection{Subject Systems}
\label{subsec:subject_sys}

To perform the preliminary evaluation we selected two current open source projects Rhino and Mylyn.Tasks as our subject systems. Rhino is a Java engine for JavaScript language, and Mylyn.Tasks is a sub project of the broader Mylyn framework. The projects were previously studied by other researchers when evaluating FLTs \cite{Zamani2014,Kevic2014}. Both projects fall into a category of small to medium sized software systems, they conveniently use Git and \emph{Bugzilla}\footnote{\url{https://www.bugzilla.org/}}, and have significant proportions of Java source code. We locked the projects at a given revision as shown in Table~\ref{tbl:subject_stats} and employed our custom built tools and \emph{cloc}\footnote{\url{http://cloc.sourceforge.net/}} utility to collect interesting Java source code statistics.

As seen from Table~\ref{tbl:subject_stats} for each level of granularity we distinguish two cases, when only the most recent or all changesets are considered. The average number of distinct commits per artifact are comparable for each granularity level except when file level granularity including the most recent commits only was taken (i.e. 9.44 compared to 11.48 for Rhino and Mylyn.Tasks respectively). Unsurprisingly, the method level granularity with the most recent changesets shows the lowest average number of distinct changesets for the observed projects, while the file level granularity where all changesets are retrieved shows the highest average number of distinct changesets.

\subsection{Metrics and Method}
\label{subsec:metrics_method}
To reasonably answer the research questions we need a set of sound quantitative measures. Textual FLTs utilizing IR rely on several frequently employed measures: \emph{effectiveness}, \emph{mean average precision} (MAP), and \emph{mean reciprocal rank} (MRR).

\begin{itemize}
\item The effectiveness of FLT is essentially the position of the first relevant document in the ranked list of results \cite{Dit2012a,Poshyvanyk2007b}. The ranked list is generally shown in descending relevance order starting from the first position. Therefore the closer a document is ranked to the beginning of the list, the more relevant it is. Since many FLTs are associated with finding an entry point of a feature (i.e. any program artifact constituting a feature), we consider the effectiveness as the most important metric when evaluating the FLT and will use it when answering to \textbf{RQ1}, \textbf{RQ2}, and \textbf{RQ3}. To answer \textbf{RQ1} we additionally use the MRR and the MAP metrics. 
\item The MRR takes an inverse of each of the effectiveness metric values over a set of queries and returns their mean\cite{Baeza-Yates1999}. Higher values of this measure signal a more efficient technique. The MRR conveniently describes the IR techniques when only the highest correct document positions are considered. 
\item The MAP considers the average of precisions of all relevant documents at their positions for a set of queries \cite{Baeza-Yates1999}. The larger the amount of correctly predicted answers and the closer they are to the beginning of the ranked list, the more efficient is the technique. Higher values of this metric will indicate more efficient technique. The measure describes how well a technique performs at locating all relevant documents.
\end{itemize}

Before we can apply IR measures to different settings of the technique, the correct locations of the features in source code have to be discovered. One way of establishing such links between a feature and its source code is to call for system experts or to manually explore the program of interest and designate relevant pieces of code to the features \cite{Ratanotayanon2010,Chen2001}. However, domain experts are not always available and manual location of features by a non expert could lead to potentially erroneous results. A reasonable alternative is to \emph{replay} resolved change requests, where results are already known. Such a \emph{reenactment} approach was found useful by many FLTs' researchers \cite{Poshyvanyk2012,Abebe2011,Petrenko2013}. In our case, we first select a change request marked either ``FIXED'' or ``RESOLVED'' for which a changeset could be traced. This change request and changeset serves as a threshold so that all the changesets that appear before that changeset will be utilized by ACIR to build and index the search corpus. The changesets that appear after the threshold will be used for the evaluation of the technique using a reenactment approach. The request descriptions are extracted to emulate user queries and relevant changesets point to locations in source code. Thus, we were able to identify 19 matching change requests in Rhino and 20 such change requests in Mylyn.Tasks. We found that the most reliable way of binding change requests with appropriate changesets is to scan the entire changeset history and find those changesets that explicitly mention change request's number.

Finally, to answer \textbf{RQ2} we need a sound method to compare the effort at different levels of granularity. Since FL is concerned with finding the first most relevant artifact, we have chosen effectiveness as our primary measure of effort. The effectiveness will convey the position of the very first relevant document discovered. Therefore, when software artifacts' level of granularity matches that of a search intent, the effort is equal to the effectiveness. However, the difference in granularity will imply an adjustment to this measure has to be made before the results of varying granularity levels can be compared. Petrenko and Rajlich describe the effort as the amount of software artifacts a developer has to inspect before reaching the correct artifact\cite{Petrenko2013}. Hence we adapted and augmented their approach to deal with 2 separate cases. In the first scenario (Case I), a user is searching for artifacts of file level granularity, while the results are presented at method level. Then we assume that the first method entry leading to a correct file should be considered the correct answer, whereas multiple method entries leading to the same incorrect file should be counted as one. Another situation (Case II) is when a user is looking for artifacts of method level granularity, while the results are returned at file level. In this case we assume that a user has to inspect all the methods in a file until he discovers the correct method. Therefore the effectiveness is then a sum of all methods of all files a user has traversed before the correct method in the correct file was found.

\section{Results and Analysis}
\label{sec:discussion}

\begin{table}[!t]
\renewcommand{\arraystretch}{1.3}
\caption{ACIR Descriptive Statistics}
\label{tbl:desc_stats}
\centering
\resizebox{\columnwidth}{!}{\begin{tabular}{cllrrrr}
\hline
\multirow{2}{*}{\begin{tabular}[c]{@{}c@{}}Subjects \\ System\end{tabular}} & \multicolumn{1}{c}{\multirow{2}{*}{Granularity}} & \multicolumn{1}{c}{\multirow{2}{*}{\begin{tabular}[c]{@{}c@{}}Commit \\ range\end{tabular}}} & \multicolumn{2}{r}{Effectiveness} & \multirow{2}{*}{MAP \%} & \multirow{2}{*}{MRR \%} \\ \cline{4-5}
                                                                            & \multicolumn{1}{c}{}                             & \multicolumn{1}{c}{}                                                                         & Median          & Mean            &                         &                         \\ \hline
\multirow{4}{*}{Rhino}                                                      & \multirow{2}{*}{Method}                          & Most recent                                                                                  & 60              & 183.43          & 7.9                     & 12.06                   \\
                                                                            &                                                  & All                                                                                          & 75              & 271.53          & 6.84                    & 11.65                   \\
                                                                            & \multirow{2}{*}{File}                            & Most recent                                                                                  & 4               & 12.22           & 39.63                   & 44.32                   \\
                                                                            &                                                  & All                                                                                          & 5               & 18.11           & 34.27                   & 43.82                   \\
\multicolumn{1}{l}{}                                                        &                                                  &                                                                                              &                 &                 &                         &                         \\
\multirow{4}{*}{Mylyn.Tasks}                                                & \multirow{2}{*}{Method}                          & Most recent                                                                                  & 225.5           & 496.63          & 10.71                   & 14.02                   \\
                                                                            &                                                  & All                                                                                          & 198             & 422.06          & 10.99                   & 18.46                   \\
                                                                            & \multirow{2}{*}{File}                            & Most recent                                                                                  & 44              & 62.26           & 15.97                   & 19.76                   \\
                                                                            &                                                  & All                                                                                          & 21              & 53.84           & 19.8                    & 21.93                   \\ \hline
\end{tabular}}
\end{table}

\begin{figure*}[!t]
\centering
\subfloat[]{\includegraphics[width=0.45\textwidth]{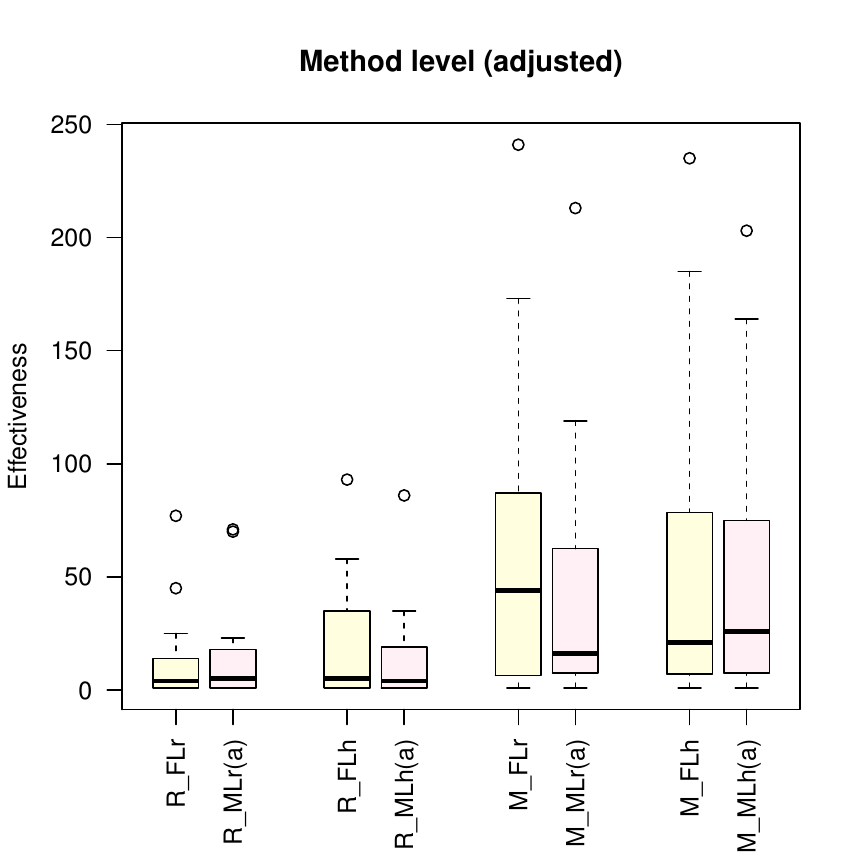}
\label{fig:eff:method_adj}}
\hfil
\subfloat[]{\includegraphics[width=0.45\textwidth]{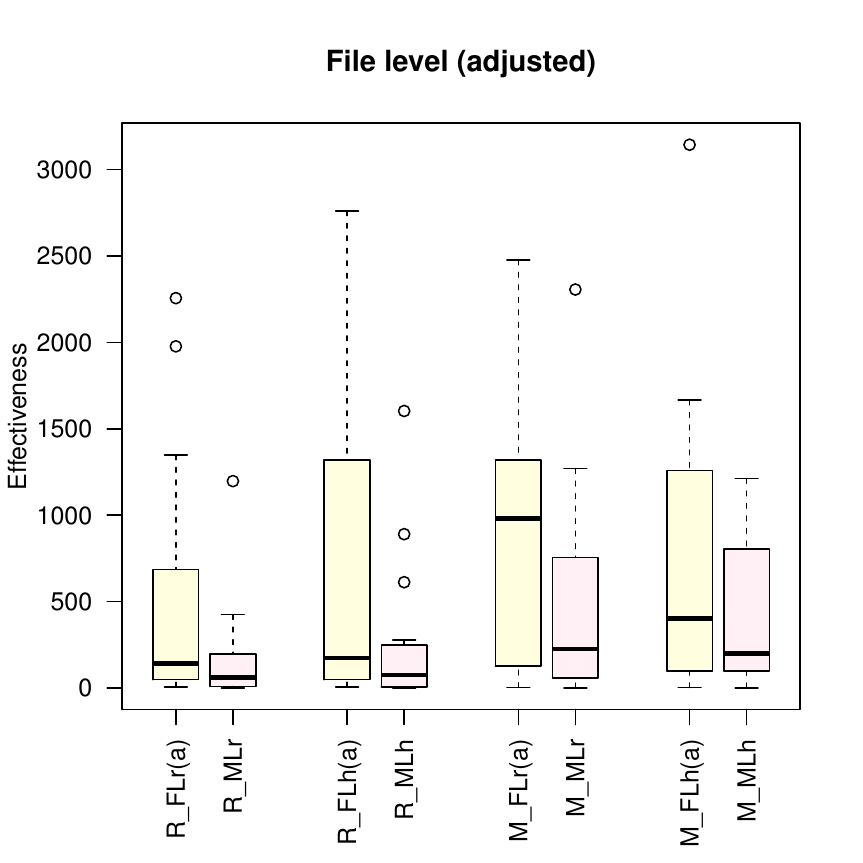}
\label{fig:eff:file_adj}}
\caption{The effort of different levels of granularity: a) when the method level effectiveness is adjusted; b) when the file level effectiveness is adjusted. \emph{* The legend to read the data is as in Table~\ref{tbl:effort_stats}. Additionally: R - Rhino, M - Mylyn.Tasks.}}
\label{fig:boxplot_eff}
\end{figure*}

\begin{figure*}[!t]
\centering
\subfloat[]{\includegraphics[width=0.45\textwidth]{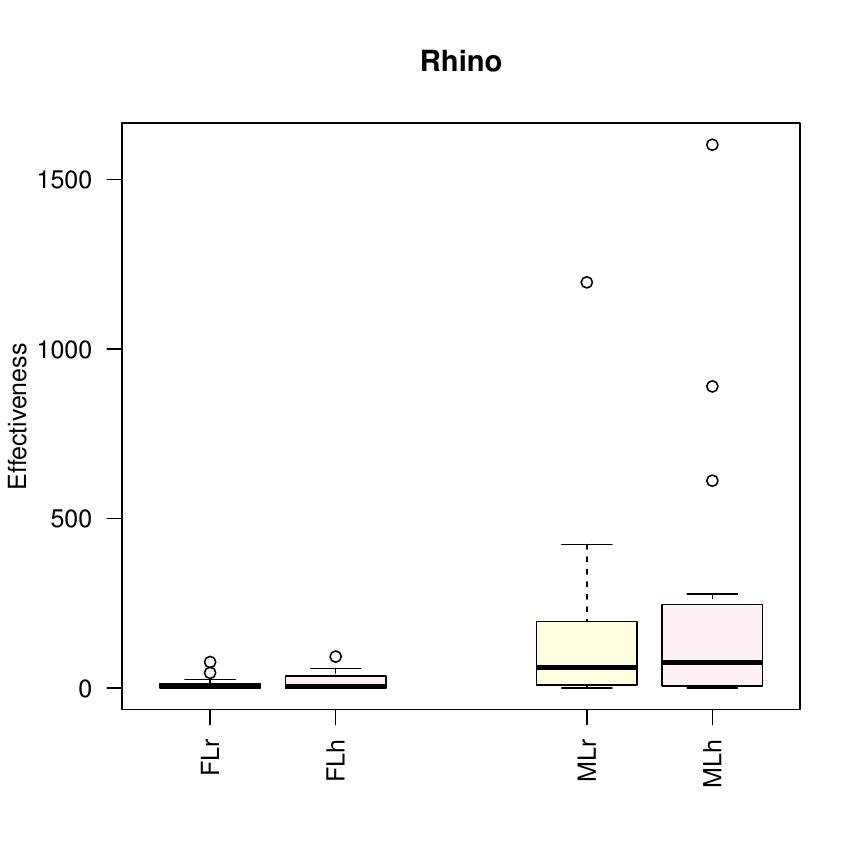}
\label{fig:range:rhino_range}}
\hfil
\subfloat[]{\includegraphics[width=0.45\textwidth]{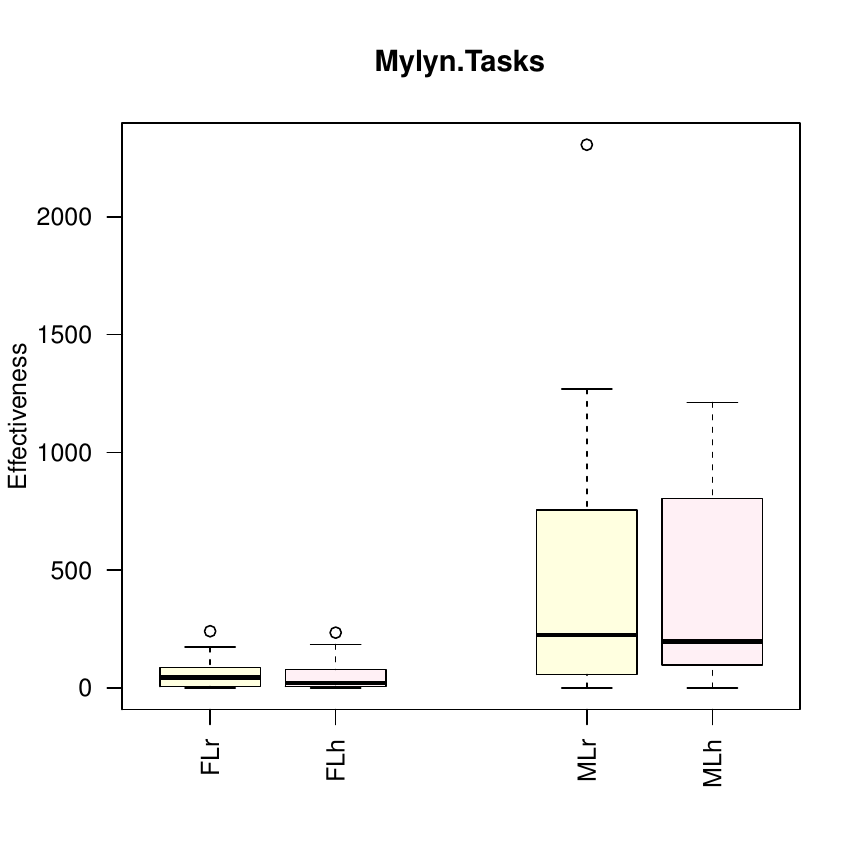}
\label{fig:range:mylyn_range}}
\caption{The effectiveness of different changeset ranges: a) in Rhino; b) in Mylyn.Tasks. \emph{* The legend to read the data is as in Table~\ref{tbl:effort_stats}.}}
\label{fig:boxplot_eff_range}
\end{figure*}

In this section we report the results of this empirical study and analyse them to answer the research questions stated in Section~\ref{sec:experiment}. In Table~\ref{tbl:desc_stats} the metrics of 4 different configurations for each subject system are presented. Table~\ref{tbl:comp_desc_stats} compares data of existing baseline IR based FLTs against ACIR. For each configuration we calculate the effort based on the adjustment scenarios discussed in Section~\ref{sec:experiment} and compile them in Table~\ref{tbl:effort_stats}. The effort values are further analysed in boxplots of Figure\ref{fig:boxplot_eff}. Table~\ref{tbl:range_stats} and the boxplots of Figure~\ref{fig:boxplot_eff_range} show the effectiveness of ACIR for different ranges of changeset inclusion.

As stated in Section~\ref{sec:experiment}, \textbf{RQ1} is: \emph{How efficient is ACIR?} To answer this question we first compare the results between Rhino and Mylyn.Tasks and then compare against existing FLTs. When comparing subject systems, the technique showed more efficient results when applied to the Rhino sample set (see Table~\ref{tbl:desc_stats}). For all appropriate settings (i.e. of similar granularity level and changeset range) the effectiveness was better for Rhino project. In one significant case (i.e. the technique configured at a file level of granularity and including the most recent changesets) the technique resulted in the median of effectiveness 10 times better than that of Mylyn.Tasks. At the method level of granularity though, Mylyn.Tasks showed MAP and MRR to be higher by 60\% and 58\% respectively in cases when all changesets were included. However, at a file level of granularity MAP and MRR of Rhino were almost 2 times higher than those of Mylyn.Tasks. The effectiveness, the MAP, and the MRR are poorer for both systems at a method level granularity. This is because documents of a file level granularity will likely have more matching terms against a search query and/or their term scores boosted, which in turn will increase their cosine similarity.

To compare against existing FLTs we gathered statistics of appropriate approaches reported in the FL literature and compiled them into Table~\ref{tbl:comp_desc_stats}. We applied rigid criteria when selecting these approaches: they had to employ comments and identifiers as their data source, they had to use algebraic IR model such as VSM or LSI, their scoring function had to use TF*IDF formula, their granularity had to match file or method level, their subject systems had to be written in Java, and they had to report comparable metrics such as effectiveness MAP or MRR. We found 6 studies that report on such approaches\cite{Zamani2014,Dit2012a,Dit2011b,Sisman2013,Petrenko2013,Scanniello2011}. We draw this selection from a population of approximately 34 research papers identified during our literature review of textual FLTs that include empirical studies and that were carried out between 2011 and 2015. Of those 6 approaches, 3 used Rhino as a subject system\cite{Zamani2014,Dit2012a,Dit2011b}, whereas the rest reported on other Java projects\cite{Sisman2013,Petrenko2013,Scanniello2011}. We consider those approaches that studied Rhino to be slightly more relevant in terms of comparison. In Table~\ref{tbl:comp_desc_stats} we report parameters of each approach only if authors state them explicitly, otherwise we assume baseline IR using VSM and TF*IDF function is employed. In those cases when we had to rely on the boxplot data \cite{Dit2012a,Dit2011b}, the metrics are presented as a range of values. Along with each metric of all these 6 approaches we report corresponding metrics of ACIR derived from the data shown in Table~\ref{tbl:desc_stats} and highlight them in bold if they show better results. The first number inside the parentheses is the average of the metric at a given granularity and the second number is the value of the best performing configuration at that same granularity. In case of non matching subject systems we derive the average of both Rhino and Mylyn appropriate metrics. As can be seen from Table~\ref{tbl:comp_desc_stats} the median of the effectiveness of ACIR was better in 4 cases out of 5. The mean of the effectiveness was better in 2 cases out of 5 when the average effectiveness was taken and in 3 cases out 5 when the effectiveness of the best configuration was taken. The MAP and the MRR of ACIR were better for all reported approaches (n.b. in some cases the MAP was significantly better - see Zamani'14 in Table~\ref{tbl:comp_desc_stats}). This preliminary data on this small dataset suggests, that the technique is competitive and shows efficient performance even in its very basic design (i.e. using standard VSM, TF*IDF, and no advanced preprocessing of data except for Porter stemming).

\begin{table*}[!t]
\renewcommand{\arraystretch}{1.3}
\caption{Comparing ACIR with Previous Work}
\label{tbl:comp_desc_stats}
\centering
\resizebox{\textwidth}{!}{\begin{tabular}{lllllrrrr}
\hline
\multirow{2}{*}{Technique}           & \multirow{2}{*}{\begin{tabular}[c]{@{}l@{}}Subject\\ System\end{tabular}}             & \multirow{2}{*}{Granularity} & \multirow{2}{*}{\begin{tabular}[c]{@{}l@{}}IR\\ model\end{tabular}} & \multirow{2}{*}{\begin{tabular}[c]{@{}l@{}}Score\\ function\end{tabular}} & \multicolumn{2}{c}{Effectiveness}                                                                                               & \multirow{2}{*}{\begin{tabular}[c]{@{}r@{}}MAP\\ \%\end{tabular}}      & \multirow{2}{*}{\begin{tabular}[c]{@{}r@{}}MRR\\ \%\end{tabular}} \\ \cline{6-7}
                                     &                                                                                       &                              &                                                                     &                                                                           & Median                                                        & Mean                                                            &                                                                        &                                                                   \\ \hline
Matching subject systems             &                                                                                       &                              &                                                                     &                                                                           &                                                               &                                                                 &                                                                        &                                                                   \\ \cline{1-1}
Zamani'14\cite{Zamani2014}         & Rhino                                                                                 & File                         & VSM                                                                 & TF*IDF                                                                    & \begin{tabular}[c]{@{}r@{}}5 \\ (\textbf{4.5; 4})\end{tabular}         & \begin{tabular}[c]{@{}r@{}}8.6 \\ (15.17; 12.22)\end{tabular}   & \begin{tabular}[c]{@{}r@{}}2.89\\ (\textbf{36.95; 39.63})\end{tabular}          & \begin{tabular}[c]{@{}r@{}}38.0\\ (\textbf{44.07; 44.32})\end{tabular}     \\ \cline{2-2}
Dit'12\cite{Dit2012a}              & Rhino                                                                                 & Method                       & LSI                                                                 &                                                                           & \begin{tabular}[c]{@{}r@{}}70-100 \\ (\textbf{67.5; 60})\end{tabular}  & \begin{tabular}[c]{@{}r@{}}200 \\ (227.48; \textbf{183})\end{tabular}    &                                                                        &                                                                   \\ \cline{2-2}
Dit'11\cite{Dit2011b}              & Rhino                                                                                 & Method                       & LSI                                                                 &                                                                           & \begin{tabular}[c]{@{}r@{}}100-120\\  (\textbf{67.5; 60})\end{tabular} & \begin{tabular}[c]{@{}r@{}}300-350\\ (\textbf{227.48; 183})\end{tabular} &                                                                        &                                                                   \\
                                     &                                                                                       &                              &                                                                     &                                                                           &                                                               &                                                                 &                                                                        &                                                                   \\
\multicolumn{2}{l}{Non matching subject systems}                                                                             &                              &                                                                     &                                                                           &                                                               &                                                                 &                                                                        &                                                                   \\ \cline{1-2}
Sisman'13\cite{Sisman2013}         & \begin{tabular}[c]{@{}l@{}}Eclipse\\ Chrome\end{tabular}                              & File                         &                                                                     & TF*IDF                                                                    &                                                               &                                                                 & \begin{tabular}[c]{@{}r@{}}20.89\\ 15.35\\ (\textbf{27.42; 29.72})\end{tabular} &                                                                   \\ \cline{2-2}
Scanniello'11\cite{Scanniello2011} & \begin{tabular}[c]{@{}l@{}}jEdit\\ Eclipse\\ ATunes\\ Art of Illusion\end{tabular}    & Method                       & VSM                                                                 & TF*IDF                                                                    & \begin{tabular}[c]{@{}r@{}}217\\ (\textbf{139.63; 129})\end{tabular}   & \begin{tabular}[c]{@{}r@{}}630\\ (\textbf{343.41; 302.75})\end{tabular}  &                                                                        &                                                                   \\ \cline{2-2}
Petrenko'13\cite{Petrenko2013}     & \begin{tabular}[c]{@{}l@{}}Adempiere\\ DrJava\\ JabRef\\ jEdit\\ Megamek\end{tabular} & Method                       &                                                                     &                                                                           & \begin{tabular}[c]{@{}r@{}}10\\ (139.63; 129)\end{tabular}    & \begin{tabular}[c]{@{}r@{}}190\\ (343.41; 302.75)\end{tabular}  &                                                                        &                                                                   \\ \hline
\end{tabular}}
\end{table*}

\begin{table}[!t]
\renewcommand{\arraystretch}{1.3}
\caption{Effort for File vs Method Level Cases of Granularity}
\label{tbl:effort_stats}
\centering
\resizebox{\columnwidth}{!}{\begin{tabular}{llrrr}
\hline
\multicolumn{1}{c}{\begin{tabular}[c]{@{}c@{}}Subject\\ Systems\end{tabular}} & \multicolumn{1}{c}{\begin{tabular}[c]{@{}c@{}}Granularity\\ levels\end{tabular}} & \multicolumn{2}{c}{Effort} & \multicolumn{1}{c}{\begin{tabular}[c]{@{}c@{}}Effort\\ change \%\end{tabular}} \\ \hline
Case I                                                                        &                                                                                  &             &              &                                                                                \\ \cline{1-1}
\multirow{2}{*}{Rhino}                                                        & FLr vs MLr(a)                                                                    & {\bf 12.22} & 14.59        & +19.34                                                                         \\
                                                                              & FLh vs MLh(a)                                                                    & 18.11       & {\bf 13.71}  & {\bf -24.3}                                                                    \\
\multirow{2}{*}{Mylyn.Tasks}                                                  & FLr vs MLr(a)                                                                    & 62.26       & {\bf 43.05}  & {\bf -30.85}                                                                   \\
                                                                              & FLh vs MLh(a)                                                                    & 53.84       & {\bf 50.47}  & {\bf -6.26}                                                                    \\
                                                                              &                                                                                  &             &              &                                                                                \\
Case II                                                                       &                                                                                  &             &              &                                                                                \\ \cline{1-1}
\multirow{2}{*}{Rhino}                                                        & FLr(a) vs MLr                                                                    & 519.67      & {\bf 183.43} & {\bf -64.70}                                                                   \\
                                                                              & FLh(a) vs MLh                                                                    & 646.22      & {\bf 271.53} & {\bf -57.98}                                                                   \\
\multirow{2}{*}{Mylyn.Tasks}                                                  & FLr(a) vs MLr                                                                    & 905.84      & {\bf 496.63} & {\bf -45.17}                                                                   \\
                                                                              & FLh(a) vs MLh                                                                    & 752.32      & {\bf 422.06} & {\bf -43.9}                                                                    \\ \hline
\multicolumn{5}{c}{{\it \begin{tabular}[c]{@{}c@{}}The legend to read the data: FL - file level, ML - method level, \\ r - recent, h - all historical, a - adjusted\end{tabular}}}                                                                                            
\end{tabular}}
\end{table}

\begin{table}[!t]
\renewcommand{\arraystretch}{1.3}
\caption{Effectiveness for All Historical vs Recent Cases of Changeset Range}
\label{tbl:range_stats}
\centering
\resizebox{\columnwidth}{!}{\begin{tabular}{clrrr}
\hline
\multirow{3}{*}{\begin{tabular}[c]{@{}c@{}}Subject\\ Systems\end{tabular}} & \multirow{3}{*}{Granularity} & \multicolumn{2}{c}{Effectiveness}                                                             & \multicolumn{1}{c}{\multirow{3}{*}{\begin{tabular}[c]{@{}c@{}}Eff.\\ change for recent\\ commits \%\end{tabular}}} \\ \cline{3-4}
                                                                           &                              & \multicolumn{1}{l}{\multirow{2}{*}{Historical}} & \multicolumn{1}{l}{\multirow{2}{*}{Recent}} & \multicolumn{1}{c}{}                                                                                               \\
                                                                           &                              & \multicolumn{1}{l}{}                            & \multicolumn{1}{l}{}                        & \multicolumn{1}{c}{}                                                                                               \\ \hline
Rhino                                                                      & File                         & 18.11                                           & 12.22                                       & +32.52                                                                                                             \\
\multicolumn{1}{l}{}                                                       & Method                       & 271.53                                          & 183.43                                      & +32.45                                                                                                             \\
\multirow{2}{*}{Mylyn.Tasks}                                               & File                         & 53.84                                           & 62.26                                       & -15.64                                                                                                             \\
                                                                           & Method                       & 422.06                                          & 496.63                                      & -17.67                                                                                                             \\ \hline
\end{tabular}}
\end{table}

To answer \textbf{RQ2} \emph{How does granularity of ACIR affect the effort?} we need to adjust the effectiveness for two cases (Case I and Case II) as discussed in Section~\ref{subsec:metrics_method}. The \emph{observed} effectiveness statistics are significantly better when the technique is applied at file level granularity (see Table~\ref{tbl:desc_stats}). This data supports a common observation in textual FLTs where the techniques of coarser granularity demonstrate better effectiveness values. Based on the effectiveness statistic we measure the effort according to Section~\ref{subsec:metrics_method} and present the results in Table~\ref{tbl:effort_stats}. According to the data in the table, the technique at method level of granularity reduces the effort by up to 31\% for Case I and up to 65\% for Case II. In one case \emph{FLr vs MLr(a)} the effort was increased by 19\%, though the difference in absolute numbers was not that significant (i.e. 12.22 vs 14.59 respectively). We then take a closer look at the effort data in Figure~\ref{fig:eff:method_adj} and Figure~\ref{fig:eff:file_adj} for cases I and II respectively. As could be seen from Figure~\ref{fig:eff:method_adj} for two cases \emph{FLh-vs-MLh(a)} and \emph{FLr-vs-MLr(a)} described in Table~\ref{tbl:comp_desc_stats}) there is an obvious difference in the data distribution of boxplots (i.e. group 2 and 3) that further supports the initial observation that less effort is required when the technique is configured at a method level of granularity. The difference is not that obvious for the last group of boxplots (\emph{FLh vs MLh(a)}). Still  the \emph{interquartile range} (IQR), the upper quartile, and the upper whisker were slightly lower at method level of granularity. For all the cases in Figure~\ref{fig:eff:file_adj} there was an obvious difference in the boxplot data. Summarizing, there is a high support in this preliminary data-set for decreased effort when the technique is configured at method level granularity for at least 6 cases shown in Table~\ref{tbl:effort_stats}. One possible explanation of the observable effect could come from the statistics of Table~\ref{tbl:subject_stats}. The artifacts of a method level are described by at least 2 changesets on average, therefore potentially increasing the chance of matches between a search query and a document and thereby resulting in better effectiveness. On the other hand, there are almost 8 and 13 methods per file on the average for Mylyn.Tasks and Rhino respectively that add to the effort required to reach the artifacts at a method level. Surely, this finding depends on the definition of the effort that we have defined in Section~\ref{subsec:metrics_method}. Therefore, the answer to the \textbf{RQ2}, based on this initial study and our effort definition, is that the observed subject systems suggest that the technique configured at method level granularity decreases effort.

To answer \textbf{RQ3} \emph{How does including only the most recent, as opposed to all changesets, affects the effectiveness of ACIR?} we considered the impact of changeset range inclusion. The average effectiveness of all configurations of Rhino system reported in Table~\ref{tbl:desc_stats} shows that the inclusion of only the most recent changesets produces more efficient results. However, for Mylyn.Tasks the most efficient configurations utilized all historical changesets. The results for Rhino show that, when using the most recent changesets, the effectiveness has improved (recall that lower effectiveness is better - see Section~\ref{subsec:metrics_method}) by up to 32\% as presented in Table~\ref{tbl:range_stats}, whereas including the most recent changesets in Mylyn.Tasks results in the effectiveness decreased by up to 17\%. A look at the Rhino boxplots in Figure~\ref{fig:range:rhino_range} shows that there is more obvious difference in data distribution, than in case of Mylyn.Tasks as seen in Figure~\ref{fig:range:mylyn_range}. When we increase the number of changesets describing each document, we are likely to add new terms to the \emph{term*document} matrix of VSM or/and alter the TF part of TF*IDF score of individual terms in the documents. If sufficiently great amounts of non matching terms regarding a search query are added to the matrix it will negatively affect the cosine similarity between a query vector and a document vector. In contrast, adding more matching terms will increase the cosine similarity between vectors. In our case a difference in the observable effect could be explained if we look at the development history of subject systems. Rhino is a fairly mature project tracing its origins back to 1997 (see Table~\ref{tbl:subject_stats}). During that time the vocabulary used to describe the program concepts may have considerably evolved rendering changeset descriptions of older changesets obsolete. Mylyn.Tasks is almost 8 years younger and most likely has retained more of its original concepts to date. Addressing \textbf{RQ3}, inclusion of all or the most recent changeset descriptions indeed affects the effectiveness of the technique. However, it is not the quantity of the changesets that plays an important role but their relevance to the current state of a software application. Including only the most recent changesets will likely increase effectiveness of ACIR when the subject system is older and old changesets are less relevant.

In summary, the preliminary findings presented here suggest that ACIR is efficient and comparable to existing IR based FLTs. For the two observed subject systems we found, that ACIR configured at a method level of granularity allows an effort reduction of up to 64\%. The inclusion of older changesets of older systems seems to negatively affect the effectiveness of ACIR.

\section{Conclusion}
\label{sec:conclusion}

Textual FLTs usually employ IR and utilize comments and identifiers to describe source code artifacts. In this paper we propose a FLT called ACIR that explores fitness of VCSs' changeset descriptions as an alternative data source for FL employing IR. The technique allows users to partition source code into artifacts of file level and method level granularity and describe them collectively using all or the most recent changeset descriptions. We were interested in how efficient ACIR is when compared to other existing textual FLTs. Also we were interested in how the level of granularity and changeset range will affect the effort and the effectiveness of ACIR. To answer these questions we implemented ACIR as a Java tool and instrumented it to operate in several interesting configurations. We then selected 19 sample FL test cases from Rhino and 20 from Mylyn.Tasks and measured the performance of ACIR in each configuration as part of the empirical study. Initial results of this preliminary study suggest that ACIR's efficiency is comparable to other textual FL approaches. We also observed that method level granularity decreased search effort by up to 64\%. We found an impact of changeset range inclusion on the final results, however the impact itself largely depends on the evolution of a given software system.

There are several issues that could affect the validity of the current study. In this observational study we worked with 2 subject systems and 39 ITS change requests. According to several studies this sample size is too small for statistically significant testing and therefore future work to scale up this study should be undertaken\cite{Oates2005, Smucker2007}.

This approach relies on sound changeset descriptions and will experience decreased performance when such descriptions are absent. Further, the vocabulary of software artifacts might get polluted, when one changeset is associated with many bugs or when a changeset is associated with large management task affecting many software artifacts.

Our technique builds a corpus of documents from software artifacts of a given source code snapshot. This could affect the empirical study since software artifacts might get deleted or moved and renamed during the evolution of a software system, which happens after the selected partitioning snapshot.
 
Possible future work includes expanding this empirical evaluation to a larger data-set and, if the results support these initial findings, checking system evolution to select the changesets for inclusion into the dataset of an artifact. It would be also interesting to try to determine a method of detecting if a one-to-many changeset:bug relationship was in a system and so might limit ACIR's applicability to that system. The overall performance results suggest that ACIR has potential and should be further compared to similar IR based FLTs utilizing comments and identifiers as well as in combination with them. ACIR might further benefit from utilizing more sophisticated IR models, advanced text preprocessing, and combination with other FL approaches.

\section*{Acknowledgement}

This work was supported, in part, by Science Foundation Ireland Grants 12/IP/1351 and 10/CE/I1855 to Lero – the Irish Software Engineering Research Centre (\url{http://www.lero.ie}).



\bibliographystyle{IEEEtran}
\bibliography{IEEEabrv,library}
%



\end{document}